\documentclass[numberedappendix]{emulateapj}
\usepackage{apjfonts}
\usepackage{graphicx}
\usepackage{natbib}
\usepackage[usenames,dvips]{color}
\usepackage{amssymb, amsmath, amsbsy, epsfig, epsf}

\newcommand{\hb}{H\ensuremath{\beta}}
\def\ha{H$\alpha$}
\def\feii{\ion{Fe}{2}}
\def\oiii{[\ion{O}{3}]}

\def\rfe{$R_{4570}$}
\def\mbh{$M_{\rm BH}$}
\def\msun{\ensuremath{M_{\odot}}}
\def\redd{$L_{\rm bol}/L_{\rm Edd}$}

\def\sgmm{$\sigma_{\rm m}$}
\def\rs{$r_{\rm s}$}

\def\kmps{${\rm km\,s^{-1}}$}

\slugcomment{accepted for publication in ApJ Letter}
\shorttitle{optical/UV variability}
\shortauthors{Y. L. Ai, et al.}

\begin{document}

\title{Dependence of the optical/UV variability on the emission line
properties and Eddington ratio in active galactic nuclei}

\author{
Y. L. Ai\altaffilmark{1,2,3},
W. Yuan\altaffilmark{1,2},
H. Y. Zhou\altaffilmark{4,5},
T. G. Wang\altaffilmark{4,5},
X.-B. Dong\altaffilmark{4,5},
J. G. Wang\altaffilmark{1,2,3},
H. L. Lu\altaffilmark{4,5}}

\altaffiltext{1}{National Astronomical Observatories/Yunnan
Observatory, Chinese Academy of Sciences, Kunming, Yunnan, P.O. BOX
110, P.R.China}

\altaffiltext{2}{Key Laboratory for the Structure and Evolution of Celestial
Objects, Chinese Academy of Sciences}

\altaffiltext{3}{Graduate School of the Chinese Academy of Sciences,
19A Yuquan Road, P.O. Box 3908, Beijing 100039, China}

\altaffiltext{4}{Center for Astrophysics, University of Science and
Technology of China, Hefei, Anhui, 230026, P.R.China}

\altaffiltext{5}{Joint Institute of Galaxies and Cosmology, SHAO and
USTC}

\email{ayl@ynao.ac.cn, wmy@ynao.ac.cn}

\begin{abstract}

The dependence of the long-term optical/UV variability on the
spectral and the fundamental physical parameters for
radio-quiet active galactic nuclei (AGNs) is investigated.
The multi-epoch repeated photometric scanning data in the Stripe-82 region
of the Sloan Digital Sky Survey (SDSS) are exploited for
two comparative AGN samples (mostly quasars)  selected therein,
a broad-line Seyfert\,1 (BLS1) type sample and
a narrow-line Seyfert\,1 (NLS1) type AGN sample within redshifts 0.3--0.8.
Their spectral parameters are derived from
the SDSS spectroscopic data.
It is found that on rest-frame timescales of several years
the NLS1-type AGNs show systematically
smaller variability compared to the BLS1-type.
In fact, the variability amplitude is found to correlate,
though only moderately, with the Eigenvector\,1 parameters,
i.e., the smaller the \hb\ linewidth, the weaker
the [O\,III] and the stronger the \feii\ emission,
the smaller the variability amplitude is.
Moreover, an interesting inverse correlation
is found between the variability and the Eddington ratio,
which is perhaps more fundamental.
The previously known dependence of the variability on luminosity is not
significant, and that on black hole mass---as claimed in recent papers
and also present in our data---fades out
when controlling for the  Eddington ratio in the correlation analysis,
though these may be
partly due to the limited ranges of luminosity and black hole mass of our samples.
Our result strongly supports that an accretion disk is likely to play a
major role in producing the opitcal/UV variability.
\end{abstract}


\setcounter{footnote}{0}
\setcounter{section}{0}

\section{Introduction}

Variation of optical/UV emission on timescales from hours to
years is one of the defining characteristics of Active Galactic Nuclei (AGNs).
Numerous previous studies of AGN/quasar ensembles have shown that
the variability amplitude
is dependent on wavelength  \citep[e.g.,][]{cutr85, di96, helf01},
luminosity \citep[e.g.,][]{hook94, give99, hawk00},
time lag \citep[e.g.,][]{de05} and redshift \citep[e.g.,][]{vand04},
and yet has a large scatter.
While it is generally believed that the optical/UV light is emitted
from an optically thick accretion disk powered by a central massive black hole,
the physical processes that produce its  variability as observed
are not understood yet.
Interestingly, recent progresses link the variability with some
fundamental parameters of AGN.
A correlation of variability with \mbh\ was reported by \citet{wold07}.
Moreover, \citet{wilh08} found correlations of
the variability with both luminosity (inversely) and \mbh, based on
which the authors speculated that variability is inversely related
to the Eddington ratio (\redd).

Prominent line emission in the optical/UV band is another defining
feature of AGNs.
A set of strong correlations
has been revealed to link the emission line properties, i.e.\
a narrow \hb\ line associated with strong optical \feii, weak \oiii\
\citep[][]{boro92}.
These correlations form the so-called  eigenvector 1 (E 1)  of AGN,
which is suggestively driven by some
underlying, more fundamental physical parameters, most likely the
Eddington ratio \citep[][]{sule00, boro02, dong09a}.

However, the  relationship between the above two AGN characteristics---the
optical/UV variability and the emission line
properties (or E1)---is poorly explored observationally so far.
The same is true for the relation between the variability and the Eddington ratio.
In this Letter, we report our investigation on this issue,
as the first result of our comprehensive studies
of AGN variability by
making use of the valuable multi-epoch repeated photometric
scanning of the Stripe-82 region in the Sloan Digital Sky Survey (SDSS).
One virtue of our study improving upon previous ones is the
inclusion of
AGNs  with optical spectra characteristic of
narrow-line Seyfert\,1 (NLS1),
in addition to typical broad-line AGNs (BLAGN)
of the broad-line Seyfert\,1 (BLS1) type,
whose variability has been extensively studied.
The NLS1-type AGNs extend to the extremes in the E1 space and
to high \redd\ values.
We use the $\Lambda$-dominated cosmology with
$H_{0} =$ 70\,\kmps Mpc$^{-1}$, $\Omega_{m} = 0.3$, and
$\Omega_{\Lambda} = 0.7$.

\section{Samples and Data Analysis}

Our comprehensive study of AGN optical/UV variability
\citep[][]{ai10} is carried out based on two comparative
BLAGN samples,
a BLS1-type sample and a NLS1-type sample,
selected from the SDSS Data Release 3
in the Stripe-82 region.
The procedure of optical spectral analysis for selecting
BLAGNs has been described
in detail in our previous papers \citep[][]{zhou06,dong08}, and the
NLS1-type sample is taken from \citet[][]{zhou06}.
The sample selection and photometric data analysis, including
photometric calibration and measurement of variability, are
described in detail in \citet{ai10},
and is only briefly summarized here.

The sample selection criteria are:
(a) located within the Stripe-82 region (RA $>$ 310$\degr$ or
$<$ 59$\degr$ and -1.25$\degr$ $<$ Dec $<$ 1.25$\degr$);
(b) redshift $z<$0.8 (H$\beta$ present in spectra) and the broad
component of \hb\ or \ha\ detected at $>10\,\sigma$;
(c) classified as `star' in {\em all} of the five bands by the SDSS
photometric pipeline so as to eliminate the contamination
from host galaxy starlight;
(d) non-radio-loud, so as to eliminate possible contamination from jet emission
\citep[][]{yuan08};
(e) a BLS1- and NLS1-type dividing line as the broad H$\alpha$/H$\beta$
FWHM $= 2,200$\,\kmps, following
\citet[][see also Gelbord et al.\ 2009]{zhou06}.
All of our NLS1-type objects also
meet the conventional \oiii/\hb$<3$ criterion for
NLS1s \citep[see][for details]{zhou06}.

\begin{figure}
\centering\includegraphics[scale=0.65]{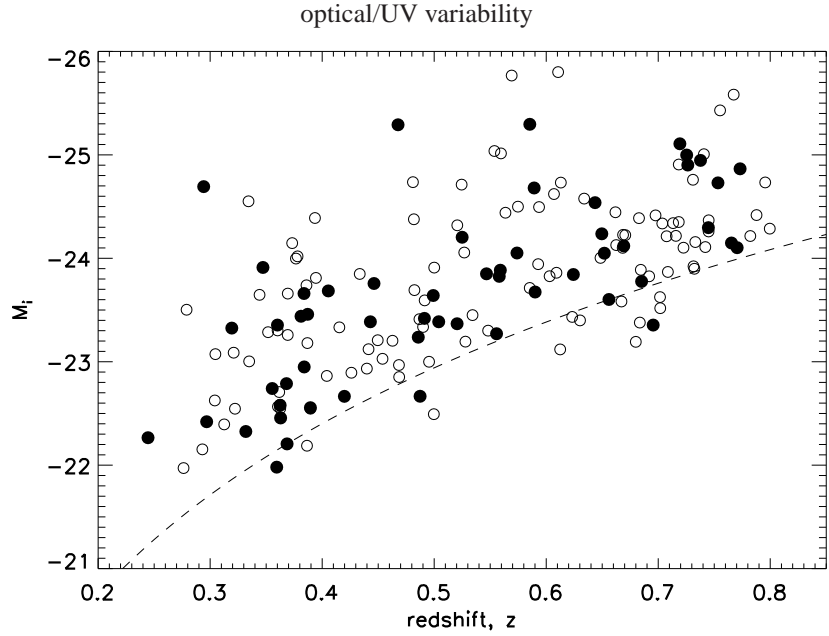}
\caption{Redshift and $i$-band luminosity distribution of our
NLS1-type AGN (filled circles) and BLS1-type (open circles) samples. The dashed curve
represents the limiting magnitude for optically-selected quasars
in the SDSS ($m_{i}=19.1$).
\label{Lum_Z}}
\end{figure}

The above selections result in 58 NLS1-type and 217 BLS1-type AGNs.
Furthermore, since we aim at a comparative study of the BLS1- and NLS1-type,
we require the two samples to have statistically compatible
distributions on the redshift--luminosity ($z-M_{\rm i}$) plane.
As the BLS1-type AGNs largely outnumber the NLS1-type ones,
a subsample of the former is then extracted
to mimic the $z-M_{\rm i}$ distribution of the latter.
We try to retain the NLS1-type sample
(with only a few outliers discarded), and prune the BLS1 sample
by randomly selecting its objects falling within sub-regions divided
on the $z-M_{\rm i}$ plane,
with a BLS1/NLS1 ratio consistent with 2:1.
This results in two final working samples,
55 NLS1- and 108 BLS1-type AGNs, which are
statistically compatible on the
redshift--luminosity plane [Figure\,\ref{Lum_Z};
the 2-D Kolmogorov-Smirnov test \citep[][]{press92}
yields a chance probability of 0.43 that they
have the same $z-M_{\rm i}$ distributions].

The two working samples consist of mostly quasars with $M_{i}<-23$\,mag and
$z\simeq$0.3--0.8.
They can be considered as optically and homogeneously selected,
with reliably measured continuum and emission line parameters
\citep[see][]{zhou06,dong08}.
Their physical parameters span almost the whole range of
BLAGNs with FWHM(H$\beta)\simeq$1000--8000\,\kmps,
black hole mass \mbh=$10^{6.5}-10^{9}$\,\msun,
luminosity ($i-$band) $M_{i}=-22\sim -26$,
and the Eddington ratio \redd=0.01--1.

The Stripe-82 region was repeatedly scanned during the SDSS-I phase
(2000-2005) under generally photometric conditions and the
data were well calibrated \citep[][]{lupt02}.
This region was also scanned repeatedly  over the
course of three 3-month campaigns in successive
three years in 2005--2007 known as the SDSS Supernova Survey (SN survey).
In this work we use the
photometric data obtained during the SDSS--I phase from Data Release 5
\citep[DR5,][]{adel07} and the SN survey during 2005.
We use the PSF magnitudes.

Observations in the SN survey were sometimes performed
in non-photometric conditions.
At the time when this work was started only the un-calibrated source
catalogs were available.
Thus we need to  calibrate out possible photometric `zero-point'
offsets in the SN data against
the DR5 magnitudes,
using  stars in the same fields as `standards'.
The photometric calibration is performed field-by-field
for each of the SN survey fields (100\,arcmin$^{2}$ patches)
in which our sample objects locate.
`Standard stars' with good measurements (high quality photometry)
are selected following the
recommendations of the
SDSS instructions\footnote{http://www.sdss.org/dr7/products/catalogs/flags.html}.
The differential magnitudes of the calibrating stars (50$\sim$200 in numbers)
between the SN survey and DR5
observations ($\Delta m=m_{\rm SN}- m_{\rm DR5}$) are
calculated, and their weighted mean is set to zero,
$\langle \Delta m \rangle=0$.
In this way the zero-point offsets of the SN survey photometry are determined,
which have (systematic) uncertainties $<$0.01\,mag
{\em relative to the reference DR5 data} for all the fields.
The total systematic errors of the calibrated SN survey photometry
are thus the quadratic sum of the above item and those of the DR5 photometry
\citep[][]{Ivez04}.
The overall (systematic and statistical) photometric errors
of the calibrated SN survey magnitudes
have a median of $\approx0.03$\,mag for the $g$, $r$, and $i$ bands, and
$\approx 0.04$\,mag for the $u$ and $z$ bands,
which are comparable to those in the DR5 data.
The reliability of our photometric calibration
can be demonstrated by examining the calibrated  SN survey data
of the 14 Landolt photometric
standard stars \citep[][]{land92} locating in Stripe-82;
none  of them is found to show detectable variability.

\begin{figure}
\centering\includegraphics[scale=0.65]{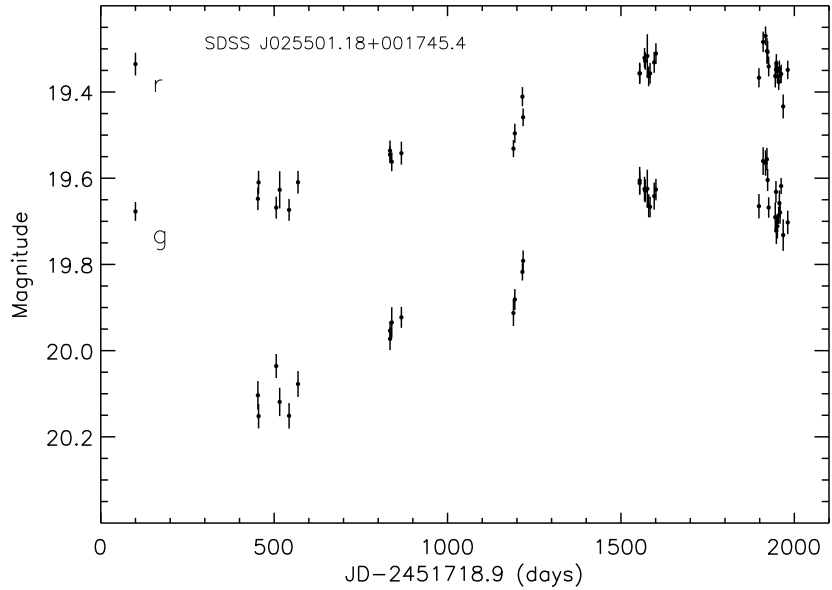}
\caption{Example light curves of the NLS1-type AGN, SDSS J025501.18+001745.4,
in the $r$ and $g$ bands.
\label{lightcurve}}
\end{figure}

For each of the objects there are typically  $\sim$27 observations
(14 from the SN survey during 2005)
spanning $\sim$5 years (see Figure\,\ref{lightcurve} for an example
lightcurve). The optical variability is commonly measured by the
variance of observed magnitudes, with the contribution due to
measurement errors subtracted; the intrinsic amplitude $\sigma_{\rm
m}$ is given by the square root of this variance
\citep[e.g.][]{vand04,sesa07}. We adopt a formalism similar to that
used in \citet[][]{sesa07},
\begin{equation}\label{eq2}
\Sigma = \sqrt{\frac{1}{n-1} \sum\limits_{i=1}^{N}(m_{i} - \langle m
\rangle)^2}  ~~,
\end{equation}
where $\langle m \rangle$ is the weighted mean, and
\begin{equation}\label{eq4}
\sigma_{\rm m}=\left\{%
\begin{array}{ll}
    (\Sigma^2-\xi^2)^{1/2}, & {\rm if~\Sigma > \xi,}\\
    0, & {\rm otherwise} .\\
\end{array}%
\right.
\end{equation}
Here the contribution to the variance due to measurement errors  $\xi$
is estimated directly from the errors of observed individual magnitudes
$\xi_{i}$
(including both the statistical and systematic errors),
as
\begin{equation}\label{eq3}
\xi^2 = \frac{1}{N} \sum\limits_{i=1}^{N} \xi_{i}^{2},
\end{equation}
rather than that in \citet[][]{sesa07} where a fitted error relation
from a large sample was used.
Such an estimation of $\xi$ was also
used in \citet{rodr97}, though in the flux rather than magnitude domain.

We approximate the derived amplitude \sgmm\ for a filter band as
that at its effective wavelength $\lambda_{0}$.
For an object of redshift $z$, $\lambda_{0}$ corresponds to a wavelength
$\lambda=\lambda_{0}/(1+z)$ in the object's rest frame.
Thus for each object the variability \sgmm\
at five rest wavelengths are sampled which correspond to the five SDSS bands.

\section{Results}

\begin{figure}
\centering \includegraphics[scale=0.85]{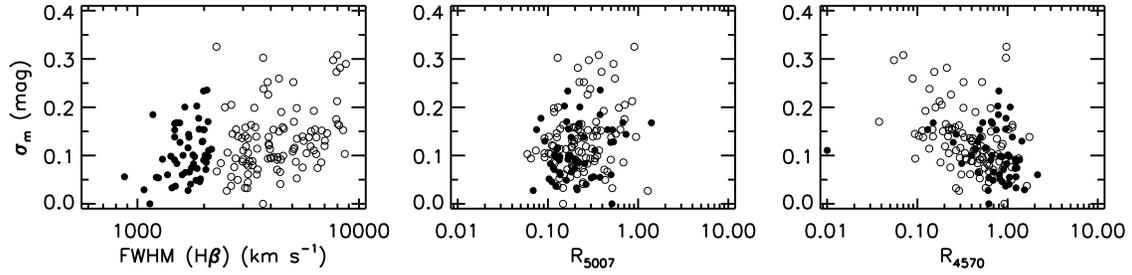}
\caption{Variability amplitude versus \hb\ linewidth (left), the
relative strength of \oiii\ (center) and \feii\ emission (right) in
{\em one of} the rest wavelength bins (2500--3300\,\AA),
as an example, for BLS1-type (open) and NLS1-type AGNs
(filled). \label{correlation_E1}}
\end{figure}

We find that the long-term variability with amplitudes \sgmm\
greater than $\approx$0.05\,mag (approximately the minimum amplitude
detectable in this work) is ubiquitous in our AGNs.
The previously known strong dependence of the amplitude on (rest-frame)
wavelength is also detected. To eliminate this
effect in the following analysis, we divide the whole rest
wavelength range sampled here, 1900--7100\,\AA, into five bins of
equal bin-size in $\log\lambda$ (see Table\,\ref{correlation}), and
treat the variability in  each bin separately. In a few objects
where more than one measurement falling into one wavelength bin, the
averaged \sgmm\ values is used for that bin.
We find that, although the wavelength dependence is negligible within a
wavelength bin, large scatters of variability clearly remain (see
Figure\,\ref{correlation_E1}).
Of particular interest, we find that
NLS1-type AGNs have systematically lower variability than BLS1-type
AGNs in all the five wavelength bins, which are moderately significant
(using the Student's {\it t}-test for the sample means
yields probability levels of 0.001--0.01).
A question is: what causes the large diversity of the
variability, or what controls the variability amplitude? It is well
known that large scatters in some of the AGN observables can be
reduced when the set of Eigenvector\,1 correlations  are considered,
in which AGNs lie continuously  with NLS1-type clustering at
one end. Below we extend these correlations to including variability
amplitude.

\subsection{Correlations with the E1 parameters}

Correlations are tested between the variability amplitude \sgmm\
and several well known E1 parameters, namely,
\hb\ FWHM, the relative strength of the \oiii\ and \feii\ emission lines.
The latter two are measured by the intensity ratios to \hb,
i.e. $R_{5007} \equiv $\rm \oiii $\lambda 5007 / \rm H\beta$
(where the total flux of \hb\ is used) and
\rfe $\equiv$\feii $\lambda \lambda 4434-4684 / \rm H\beta^b$
(where
\feii\ $\lambda\lambda4434-4684$ denotes the \feii\
multiplets flux integrated over 4434--4684\,\AA,
and H$\beta^b$ the flux of the broad component of \hb).
The non-parametric Spearman rank statistic is used.
The correlations are tested for the combined BLS1- and NLS1-type AGN sample
as well as for each sample separately.
The results are summarized in Table\,\ref{correlation},
with the Spearman's coefficient $r_{\rm s}$
and the two-tailed probability for the null hypothesis of no correlation.
Figure\,\ref{correlation_E1} shows
the data in {\em one} of the wavelength bins as examples.

For the combined AGN sample in all the wavelength bins,
there exist moderate yet interesting correlations between the
variability  and the three E1 parameters,
which are statistically significant.
The smaller the \hb\ linewidth is, the weaker the [O\,III] and the stronger
the \feii\ emission, and the smaller the variability amplitude. Among
them, the strongest correlations are those with the Fe\,II strength
$R_{4570}$ ($r_{\rm s} \sim -0.4$), followed by those with
FWHM(\hb), while those with $R_{5007}$ are the weakest.
Some of the correlations still remain significant even when the two
samples are considered separately.

It is thus clear that the large scatter in the optical/UV
variability as observed can, at least to some extent, be attributed  to the
distribution of AGN across the E1 parameter space.
Our finding of the lower ensemble variability of NLS1-
compared to BLS1-type AGNs
is just a manifestation of these correlations, for the former
have smaller linewidths by definition and generally weaker [O\,III] and
stronger \feii\ emission than the latter.
Given such significant and interesting correlations, we suggest
that the E1 correlations may be extended to those involving the
variability in the optical/UV band.

\subsection{Link with the Eddington ratio, luminosity and black hole mass}

There have been suggestions that the underlying driver of the E1
correlations is most likely  the Eddington ratio
\citep[][]{boro92, sule00, boro02}.
Here we test the dependence of the  variability on \redd\  explicitly.
We also examine the variability dependence on luminosity and \mbh,
as claimed in previous studies.
\mbh\ are estimated using the scaling relation given in \citet{gree05} from
the broad \hb\ FWHM and the
5100\,\AA\ luminosity ($L_{5100}$), which are taken
from or measured in the same way as in \citet{zhou06}.
To calculate \redd\ the
bolometric luminosities are estimated as $L_{\rm bol}=9L_{5100}$ \citep{elvi94}.
The results of the correlation tests are
listed in Table\,\ref{correlation}, and the data in the
2500-3300\,\AA\ bin are demonstrated
in Figure\,\ref{correlation_mass_edd} as an example.

Of particular interest, the strongest correlation is
an inverse correlation of the variability with \redd\
(\rs$\sim-0.4$), which is the case for all the wavelength bins. The
correlation is present in not only the combined  BLS1- and NLS1-type AGN sample
but also the individual samples separately, although it is only
marginal in the NLS1-type sample alone
(possibly due to its narrow \redd\ range).
These correlations are as strong as the above
\sgmm--$R_{4570}$ correlation.
In addition,
the combined two samples also show a positive correlation between the
variability and black hole mass, though it is considerably weaker
compared to that with \redd.
For luminosity ($L_{5100}$ used here),
however, there is essentially no significant correlation
found in all the wavelength bins
except 2500--3300\,\AA, in which  a weak inverse correlation
may be marginally present ($P\sim0.01$).

Since \redd\ appears to correlate with \mbh\ in our samples
(as commonly found in optically selected AGN samples
with a limited luminosity range),
one or both of their correlations with the variability may not be as
significant as it appears to be, or one may even be spurious and
induced  from the other. We try to eliminate this possible effect of
the third variant using the partial Spearman correlation test
(see the last two columns in Table\,\ref{correlation} for results).
We first re-examine the correlation between \sgmm\ and \mbh\ by taking  into
account their respective correlations with \redd. It turns out that,
in all the wavelength bins, the correlations with \mbh\ vanish when
the dependence on \redd\ is taken into account.
However, the same treatment for the \sgmm--\redd\
relation, with the effect of \mbh\ eliminated, still yields
significant, though somewhat reduced, inverse correlations in
three wavelength bins. We thus conclude that the \sgmm--\redd\
correlation is genuine, regardless of the \sgmm--\mbh\ correlation.
The latter, as apparently seen in the data and previously reported
\citep[][]{wold07, wilh08}, is consistent with being induced from
the former.

\section{Discussion and conclusions}

It is intriguing that the optical/UV variability is closely related to
the AGN E1 parameters.
This can be understood in light of the same interesting correlations
with the Eddington ratio, also found here,
provided that \redd\ is indeed the underlying driver of the E1 correlations.
Since the variability--\redd\ correlation is
significant enough for the BLS1-type sample alone without including NLS1-type,
our result does not rely on the \mbh\ (thus \redd) estimation
for NLS1-type AGNs, which is still somewhat controversial.
The fact that the correlations of the variability with the E1 parameters and \redd\
are as strong as those among the E1 parameters
and \redd\ themselves suggests that the E1 correlations may be extended to
those involving the optical/UV variability.

\begin{figure}
\centering \includegraphics[scale=0.85]{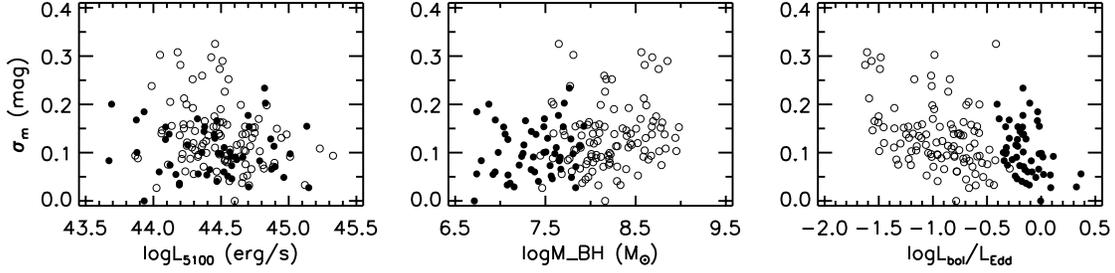} \caption{
Similar to Figure\,\ref{correlation_E1} but for
variability amplitude versus 5100\,\AA\ luminosity (left), black hole
mass (center), and the Eddington ratio (right).
\label{correlation_mass_edd}}
\end{figure}

The correlation between the variability and \mbh,
though also present in our data,
turns out to be insignificant when \redd\ is taken into account.
We thus suggest that such a correlation,
as claimed in recent studies \citep[][]{wold07, wilh08},
might be spurious and induced from the variability-\redd\ relation.
The previously known variability--luminosity inverse
correlation is found to be insignificant or at most weak in our data.
This may be
partly due to the fact that our analysis is capable of recovering
AGNs (some are NLS1-type) with small variability in the low luminosity
regime (see Figure\,\ref{correlation_mass_edd}) which were largely
missed in previous work for various reasons.
However, our morphology and redshift cuts restrict both the
luminosity and \mbh\ ranges of our samples,
which are important for comparisons with previous work.
For example, \citet{vand04} have shown that a factor of 100 change
in luminosity corresponds to a factor of 4 change in variability.
Thus it cannot be ruled out that the lack of significant dependence
on either luminosity or \mbh\ may be due to their limited ranges in our samples.

The ensemble weaker variability of NLS1- compared to BLS1-type AGNs, as
evidently found here and also suggested by \citet{klim04} based on a small
sample, is simply a natural consequence of the correlations with the
E1 parameters and the Eddington ratio. The dependence on \redd\ can
be understood qualitatively in terms of decreasing variability with
increasing radius of the accretion disk, independent of the actual
physical processes that drives the variability. Such
radius-dependent variability is suggested by the fact that the
variability amplitude increases as wavelength decreases
\citep[e.g.][]{vand04} in the context of thermal disk model. As the
accretion rate $\dot{m}$ (in units of the Eddington rate) increases,
the continuum emission region that dominates a given waveband moves
outward to a larger radius $r$ (in units of the Schwarzschild
radius) because of $\lambda\propto T^{-1}\propto (M/\dot{m})^{1/4}
r^{3/4}$; and hence decreasing variability is expected. We note that
such a trend of optical variability seems at odds with that in the
X-ray band \citep{papa04}, though in the X-ray case more convincing
results are needed. The different behavior may result from the
different mechanisms and emitting regions of the radiation, as well
as different variability mechanisms,  in the two bands
\citep[see e.g.][]{uttl06}.

Our finding of the strong correlations of the variability with the
E1 parameters and \redd\ imposes interesting constraints to AGN
variability models. Foremost, this indicates that the optical/UV
variability must be intrinsic to AGN activity; any external
scenarios, such as gravitational microlensing \citep[][]{hawk00},
multiple supernovae and star collisions \citep[][]{cid00}, can be
ruled out as the dominant process. For intrinsic variability models,
it is not clear whether the mechanism causing variability is some
minor, secondary effect or related to the main energy generation
process, as discussed in \citet{gask08}. Our result favors the
latter since the Eddington ratio is directly related to the mass
accretion rate of AGN at a given accretion efficiency.
The correlation with \redd\ as found here suggests that the accretion disk most likely
plays a critical role in producing the observed opitcal/UV
variability, since the disk structure is depending on the
mass accretion rate. For instance, the model proposed by
\citet[][]{li08} that the variability is caused by change of
accretion rates can reproduce {\em qualitatively} the observed inverse
variability--\redd\ relation (Li S., private communication).

\acknowledgments
We thank the referees for helpful comments and suggestions.
W.Y. thanks S.\ Li and B.\ Czerny for useful discussions.
This work is supported by Chinese NSF grants NSF-10533050, the National Basic
Research Program of (973 Program 2009CB824800, 2007CB815405).
Funding for the SDSS and SDSS-II was provided by the Alfred P. Sloan Foundation, the
Participating Institutions, the National Science Foundation, the U.S. Department of Energy,
the National Aeronautics and Space Administration, the Japanese Monbukagakusho,
the Max Planck Society,
and the Higher Education Funding Council for England.
The SDSS Web Site is http://www.sdss.org/.



\clearpage

\begin{deluxetable}{llllllllll}
\tabletypesize{\scriptsize} \tablewidth{0pt}
\tablecaption{Results of correlation tests for variability amplitude in rest wavelength bins:
Spearman correlation coefficients (and probability levels)
\label{correlation}} \tablehead{\colhead{Sample}  &  \colhead{$^a$Size}  &
\colhead{FWHM(\hb)}   &   \colhead{$^bR_{5007}$}
& \colhead{$^bR_{4570}$}   &   \colhead{$^b$log(L$_{5100}$)}  &   \colhead{log(\mbh)}   &   \colhead{\redd}
& $^c$\mbh  &  $^d$\redd}
\startdata
&&&\multicolumn{3}{c}{\bf 1900 $\sim$ 2500\,\AA} &&&\\
all AGNs&118(0.24) & {0.33(2e-04)} & {0.26(4e-03)} &{-0.39(1e-05)} & {-0.05(0.62)} & {0.32(4e-04)}&{-0.38(3e-05)} & {0.01(0.93)}& {-0.21(0.02)} \\
BLS1-type&81(0.28) & {0.31(5e-03)} & {0.30(6e-03)} & {-0.40(3e-04)} &{-0.08(0.47)} & {0.24(0.03)} & {-0.34(2e-03)}&-&- \\
NLS1-type&37(0.41) & {0.29(0.09)} & {0.15(0.38)} & {-0.26(0.12)} & {0.10(0.55)} & {0.27(0.12)} & {-0.21(0.22)}&-&- \\
&&&\multicolumn{3}{c}{\bf 2500 $\sim$ 3300\,\AA} &&&\\
all AGNs&158(0.20) & {0.32(3e-05)} & {0.21(7e-03)} & {-0.40(2e-07)} &{-0.16(0.04)} & {0.26(1e-03)} & {-0.38(8e-07)} &{-0.12(0.11)} & {-0.32(6e-05)}\\
BLS1-type&104(0.25) & {0.34(4e-04)} & {0.21(0.03)} & {-0.43(7e-06)} & {-0.25(0.01)} & {0.19(0.05)} & {-0.40(3e-05)}&-&- \\
NLS1-type&54(0.35) & {0.30(0.03)} & {0.18(0.18)} & {-0.30(0.03)} & {-0.08(0.57)}& {0.10(0.46)} & {-0.37(7e-03)}&-&- \\
&&&\multicolumn{3}{c}{\bf 3300 $\sim$ 4200\,\AA} &&&\\
all AGNs&159(0.20) & {0.31(6e-05)} & {0.23(3e-03)} & {-0.38(1e-06)} & {-0.13(0.09)} & {0.26(1e-03)} & {-0.37(2e-06)}& {-0.09(0.22)}& {-0.29(3e-04)}\\
BLS1-type&105(0.25) &{0.26(7e-03)}&{0.26(6e-03)}&{-0.37(9e-05)}&{-0.17(0.09)}&{0.16(0.10)}&{-0.31(1e-03)}&-&- \\
NLS1-type&54(0.35) & {0.24(0.09)} & {0.12(0.37)} & {-0.27(0.05)} & {-0.12(0.38)}& {0.04(0.76)} & {-0.35(0.01)}&-&- \\
&&&\multicolumn{3}{c}{\bf 4200 $\sim$ 5500\,\AA} &&&\\
all AGNs&163(0.20) & {0.34(1e-05)} & {0.25(1e-03)} & {-0.43(1e-08)}&{-0.08(0.33)} & {0.30(1e-04)} & {-0.38(9e-07)}&{-0.04(0.62)} & {-0.24(2e-03)}\\
BLS1-type&108(0.25) & {0.34(3e-04)} & {0.25(8e-03)} & {-0.45(1e-06)} &{-0.12(0.20)} & {0.25(9e-03)} & {-0.37(8e-05)}&-&- \\
NLS1-type&55(0.34) & {0.29(0.03)} & {0.20(0.16)} & {-0.34(0.01)} & {-0.04(0.76)}& {0.14(0.31)} & {-0.32(0.02)}&-&- \\
&&&\multicolumn{3}{c}{\bf 5500 $\sim$ 7100\,\AA} &&&\\
all AGNs&104(0.25) & {0.39(5e-05)} & {0.29(2e-03)} & {-0.43(5e-06)} &{-0.01(0.92)} & {0.35(3e-04)} & {-0.41(2e-05)}& {0.003(0.97)}& {-0.23(0.02)}\\
BLS1-type&66(0.31) & {0.44(2e-04)} & {0.34(5e-03)} & {-0.41(8e-04)} &{-0.01(0.92)} & {0.37(2e-03)} & {-0.44(2e-04)}&-&-\\
NLS1-type&38(0.41) & {0.30(0.07)} & {0.13(0.45)} & {-0.33(0.05)} & {-0.06(0.70)}& {0.13(0.44)} & {-0.36(0.03)}&-&-\\
\enddata
\tablecomments{$^a$sample size (and the critical value above
               which a correlation coefficient is found from an
                uncorrelated sample of the size at a $<$1\% probability);
                $^b$$R_{5007}$ is the intensity ratio of \oiii\ $\lambda$5007 to \hb,
               $R_{4570}$ the intensity
               ratio of the \feii\ multiplets  to \hb, $L_{5100}$ monochromatic luminosity at 5100\,\AA;
               $^c$partial correlation between \sgmm\ and \mbh, controlling for \redd;
               $^d$partial correlation between \sgmm\ and \redd, controlling for \mbh.}
\end{deluxetable}


\end{document}